\documentclass{article}
\usepackage{setspace}
\usepackage[a4paper]{geometry}
\usepackage{silence}
\usepackage{tikz}
\usetikzlibrary{arrows.meta, positioning, calc}

\usepackage[english]{babel}
\usepackage[utf8]{inputenc}
\usepackage[T1]{fontenc}

\usepackage[dvipsnames]{xcolor}

\usepackage{amsmath, amssymb, amsfonts, amsthm, mathtools}
\usepackage{mathrsfs}
\usepackage{braket}
\usepackage{cancel}
\usepackage{latexsym}

\usepackage{graphicx}
\usepackage{float}
\usepackage{subcaption}
\usepackage{listings}
\usepackage{stackengine}

\usepackage{blindtext}
\usepackage{verbatim}
\usepackage{array}
\usepackage{setspace}
\usepackage{multicol}
\usepackage{fancybox}
\usepackage{rotating}
\usepackage{enumerate}
\usepackage{textcomp}
\usepackage{sectsty}
\usepackage{comment}
\usepackage{pifont}
\usepackage{empheq}
\usepackage{titlesec}
\usepackage{abstract}
\usepackage[autostyle=true]{csquotes}

\usepackage{tikz}
\usetikzlibrary{arrows.meta,shapes}

\definecolor{micolor}{RGB}{230,240,255}
\xdefinecolor{ritsumeikan}{RGB}{217, 180, 34}
\definecolor{darkbackground}{RGB}{0, 52, 64}
\xdefinecolor{colormio}{RGB}{217, 180, 34}

\tikzset{
  mynode/.style={
    draw,
    chamfered rectangle,
    minimum width=3.5cm,
    minimum height=0.7cm,
    align=center,
    fill=micolor!50
  },
  myarrow/.style={
    -{Stealth[length=3mm]},
    thick,
    shorten >=8pt,
    shorten <=8pt
  },
  myarrowdbl/.style={
    {Stealth[length=3mm]}-{Stealth[length=3mm]},
    thick,
    shorten >=8pt,
    shorten <=8pt
  },
  pathlabel/.style={midway, sloped, fill=white, inner sep=1pt, font=\footnotesize},
  pathnum/.style={midway, sloped, circle, draw, fill=white, inner sep=1pt, font=\scriptsize}
}

\usepackage[most]{tcolorbox}
\tcbset{
  my math box/.style={
    standard jigsaw,
    colback=ritsumeikan!10, colframe=ritsumeikan, size=fbox, arc=3pt, boxrule=0.5pt,
    opacityback=0.82,
  },
  my math boxy/.style={
    standard jigsaw,
    colback=ritsumeikan!20, colframe=ritsumeikan, size=fbox, arc=3pt, boxrule=0.5pt,
    opacityback=0.82,
  }
}

\newtcbox{\mymath}[1][]{%
    nobeforeafter, math upper, tcbox raise base,
    enhanced, colframe=blue!30!black,
    colback=blue!30, boxrule=1pt,
    #1}

\titleformat{\paragraph}{\normalfont\normalsize\bfseries}{}{0pt}{}
\titlespacing*{\paragraph}{0pt}{1em}{0.7em}

\makeatletter
\renewcommand\@makefnmark{\hbox{\@textsuperscript{\normalfont\color{colormio!75!black}\@thefnmark}}}
\makeatother

\usepackage{etoolbox}
\makeatletter
\pretocmd{\tagform@}{\color{gray!50!black}}{}{}
\makeatother

\usepackage[backend=biber,style=numeric-comp,sorting=none,sortcites]{biblatex}
\usepackage{hyperref}
\hypersetup{
  colorlinks = true,
  linkcolor  = blue!80!black, 
  citecolor  = red,       
  filecolor  = red,
  urlcolor   = black
}

\usepackage[normalem]{ulem}

\begin{document}

\title{\large{\textbf{A connection between Gravitational Scalar–Tensor theories and Generalized Hybrid theories\\
}}}

\author{  \small{Jonathan Ramírez\footnote{\href{mailto:ramirez.jsg@gmail.com}{ramirez.jsg@gmail.com}}  ~and Santiago Esteban Perez Bergliaffa\footnote{\href{mailto:sepbergliaffa@gmail.com}{sepbergliaffa@gmail.com}}}\\
 \small{
 \textit{Departamento de Física Teórica, Instituto de Física,}}\\
\textit{\small{Universidade do Estado do Rio de Janeiro,}}\\
\textit{\small{Rua São Francisco Xavier 524, Maracanã,}} \\ 
\textit{\small{CEP 20550-013, Rio de Janeiro, Brazil.}}
}

\date{}
\maketitle

\begin{abstract}
{We establish a correspondence between higher–derivative gravitational scalar–tensor theories of the form $\Psi(R,(\nabla R)^2,\Box R)$ and generalized hybrid metric–Palatini models $f(R,\mathcal{R})$. Restricting to 
the physically relevant case of linear dependence on $\Box R$, we make explicit that both frameworks can be reformulated in the Einstein frame as General Relativity minimally coupled to two interacting scalar fields, thereby 
opening the possibility of finding theories that are dynamically equivalent. This correspondence provides an explicit dictionary relating the functions that define the higher–derivative theory to the hybrid function $f(R,\mathcal{R})$, allowing for reconstruction in both directions. We illustrate the usefulness of the procedure with explicit examples.}
\end{abstract}

\maketitle

\section{Introduction}
\label{sec:Introduction}
The search for consistent extensions of General Relativity (GR) has led to many alternatives, including higher-derivative and scalar--tensor theories of gravity
(see for instance \cite{Clifton2011,Berti2015}). Among the alternatives, $f(R)$ models and their generalizations, where the Lagrangian is a function of the  curvature invariants and their derivatives, provide a compelling framework to address both early- and late-time cosmological phenomena such as inflation and dark energy \cite{Amendola1993, Wands1994,
Cuzinatto2018,
Chaadaeva2024Cosmological}, as well as 
regular black holes 
\cite{Netto2023}. 
However, generic higher-derivative actions are usually plagued by Ostrogradski instabilities, introducing unphysical ghost degrees of freedom. A central challenge in the development of extended gravity models is therefore the identification of viable formulations that propagate only the healthy dynamical modes.
An important step in this direction was achieved with the construction of the so-called gravitational scalar--tensor (GST) theories, in which the action is written as a function of the Ricci scalar and its first and second order
derivatives while remaining ghost-free under suitable conditions \cite{Naruko2016}. 
Since GST theories can be considered as gravitational counterparts of multi-scalar-tensor 
(MST) theories, described by the metric and several scalar fields, it is natural to seek connections 
between GST theories and theories that can be rewritten as MST theories. 
In particular, the so-called generalized hybrid (GH) theories,
introduced in \cite{tamanini} through  a Lagrangian that is a function of the metric Ricci scalar $R$ and the Palatini Ricci scalar $\mathcal{R}$, the latter being constructed from an independent connection, can be formulated as a bi-scalar–tensor theory. 
\vspace{0.2cm}

It is important to point out that, while there are several references dealing with many aspects of the GH theories (see \cite{Gomes2025} and references therein), there are only a few applications of the GST theories in the literature
(see  
\cite{Saridakis2016, Saridakis2018,Banerjee2022}). In particular, in both cases it is notably difficult to find exact solutions of the equations of motion. 
In the present work, we shall establish an explicit correspondence between generalized higher-derivative theories of the form $f(R,(\nabla R)^2,\Box R)$ 
(introduced in
\cite{Naruko2016})
and the GH theories $f(R,\mathcal{R})$. By introducing auxiliary fields and performing conformal transformations, both frameworks can be recast into the form of Einstein's gravity minimally coupled to two scalar degrees of freedom, characterized by specific kinetic couplings and interaction potentials. {This was shown for a general $\Psi$ in \cite{Naruko2016}, and in \cite{tamanini} for the generalized hybrid theories}, 
thus establishing a
potential 
systematic dictionary between GST and GH theories, that allows to profit from the results obtained in one type of theory (such as exact solutions) in the other type.
\vspace{0.2cm}

The structure of the paper is the following. 
{In Sec.~\ref{sec:EF-Representation of GST} we review the Einstein-frame representation of GST theories, and we show that the ghost-free case can be  written as Einstein gravity minimally coupled to two scalar fields. In Sec.~\ref{sec:GH-gravity} GH gravity is introduced as well as its scalar--tensor representation in the Einstein frame. In Sec.~\ref{sec:Applications} we establish the correspondence between the two frameworks and provide illustrative applications: \textit{(i)}} the function  $f(R,\mathcal{R})$ is reconstructed starting from representative choices of $\Psi$, including purely kinetic extensions and mixed $\Psi(R,\Box R)$ models; and \textit{(ii)} 
the function $\Psi$ is obtained from known cosmological solutions in the GH scalar--tensor formulation. Finally, in Sec.~\ref{sec:Conclusions} we present a summary of our findings.

\section{\texorpdfstring{Einstein-Frame Representation of GST theories}~}
\label{sec:EF-Representation of GST}

The action that depends on the Ricci scalar and its first- and second-order derivatives, originally proposed in \cite{Naruko2016}, is given by
\footnote{With $\kappa^2 = 8\pi G$, $c=1$. 
}
\begin{equation}\label{action-Psi}
S=\frac{1}{2\kappa^2}\int d^4x\sqrt{-g}\, \Psi\left(R,(\nabla R)^2,\Box R\right),
\end{equation}

where $(\nabla R)^2\equiv\nabla^\mu R\,\nabla_\mu R$. 
In the general case, this action leads to a theory in which the degrees of freedom are those of the metric, and three scalar fields, one of the latter inevitably being a ghost
\cite{Naruko2016}.
The case in which the action depends linearly on $\Box R$ is particularly relevant, since in this setting the theory admits an Einstein-frame representation consisting of GR minimally coupled to two scalar fields 
and, as shown in \cite{Naruko2016}, is free from ghosts.
The general form of $\Psi$ for a linear dependence on $\Box R$
is
\begin{equation}\label{Psi-Naruko}
    \Psi(R,(\nabla R)^2,\Box R)
    =\mathcal{K}(R,(\nabla R)^2)
    +\mathcal{G}(R,(\nabla R)^2)\,\Box R,
\end{equation}
where $\mathcal{K}$ and $\mathcal{G}$ are arbitrary functions of $R$
and its first derivatives. Introducing a scalar $\phi$, the action takes the form
\begin{equation}\label{action-Naruko-B2}
    S=\frac{1}{2\kappa^2}\int d^4x\sqrt{-g}\,
    \Big[\mathcal{K}(\phi,(\nabla \phi)^2)
    +\mathcal{G}(\phi,(\nabla \phi)^2)\,\Box \phi
    -\lambda(\phi-R)\Big],
\end{equation}
where $\lambda$ acts as a Lagrange multiplier enforcing the constraint $\phi=R$.
This form of the action remains at most of second order in derivatives and, as we shall see, can be recast into a bi–scalar representation minimally coupled to Einstein gravity, after integration by parts and a conformal transformation.

Variation with respect to $\phi$ yields the relation
\begin{equation}
\lambda=\mathcal{K}_{\phi}+\mathcal{G}_{\phi}\Box \phi-2\nabla_\mu\left[\Big(\mathcal{K}_{X}+\mathcal{G}_{X}\Box \phi\big)\nabla^\mu\phi\right)+\Box \mathcal{G}, 
\label{eq:lambda}
\end{equation}
where $X\equiv(\nabla\phi)^2$. After integrating by parts, the action in Eq.\eqref{action-Naruko-B2}
 can be rewritten as follows:
\begin{equation}
S=\frac{1}{2\kappa^2}\int d^4x\sqrt{-g}\Big[\mathcal{K}-\lambda\phi+\lambda R-\nabla_\mu \mathcal{G}\nabla^\mu \phi\Big].
\end{equation}
By applying the conformal transformation $\hat{g}_{\mu\nu}=\lambda g_{\mu\nu}$, this action can be reformulated as a bi-scalar field theory coupled to Einstein gravity in the Einstein frame:
\begin{equation}
    S=\frac{1}{2\kappa^2} \int d^4x \sqrt{-\hat{g}}\Big[ \lambda^{-2}\mathcal{K}-\lambda^{-1}\phi + \hat{R}
  +3 \lambda^{-1}\hat{g}^{\sigma\nu} \nabla_\nu\nabla_\sigma \lambda-\frac{3}{2}\lambda^{-2}\hat{g}^{\sigma\nu}\nabla_\nu\lambda \nabla_\sigma \lambda -\lambda^{-2}\nabla_\mu \mathcal{G} \nabla^\mu \phi \Big].
\end{equation}
With the introduction of the field $\chi$ via the relation,
\begin{equation}\label{lambda}
    \lambda=e^{\sqrt{\tfrac{2}{3}}\kappa\chi},
\end{equation}
and discarding total derivatives, the action  takes the form
\begin{eqnarray}
    \nonumber
  S&=&\int d^4x\,\sqrt{-\hat g}\Bigg[
\frac{1}{2\kappa^2}\hat R
-\frac{1}{2}\hat g^{\mu\nu}\partial_\mu\chi\,\partial_\nu\chi
-\frac{1}{2\kappa^2}e^{-\sqrt{\tfrac{2}{3}}\kappa\chi}\,
\hat g^{\mu\nu}\partial_\mu\mathcal{G}(\phi,(\hat\nabla \phi)^2)\,\partial_\nu\phi+ \\ && ~~~~~~~~~~~~~~~~~~~~~~~~~~~~~~
-\frac{1}{2\kappa^2}e^{-\sqrt{\tfrac{2}{3}}\kappa\chi}\,\phi
+\frac{1}{2\kappa^2}e^{-2\sqrt{\tfrac{2}{3}}\kappa\chi}\,\mathcal{K}(\phi,(\hat\nabla \phi)^2)
\Bigg].
\label{eq:Einstein-fBB0}
\end{eqnarray}
It is seen that the kinetic sector for $\chi$ is canonical by construction. 
In order to simplify  
the kinetic term of $\phi$, 
the function $\mathcal{G}$ is assumed to depend solely on $\phi$, \emph{i.e.}, $\mathcal{G}(\phi,(\nabla \phi)^2)=\mathcal{G}_1(\phi)$ (in such a way that the resulting kinetic term is free from higher-order derivatives of $\phi$)
\footnote{As shown in \cite{Naruko2016}, this choice is equivalent to setting $\mathcal{G}=0$, by redefining the function $\mathcal{K}$, see also Eq.
\eqref{eq:Einstein-KG}. We choose to keep the contribution from $\mathcal{G}_1$ for later comparison with theories that are defined in terms of $\Box R$, without terms with $(\nabla R)^2$, see Sect.\ref{subsec:fRboxR}.}
, and 
the function $\mathcal{K}$ is 
taken as a linear function in $(\nabla R)^2$, namely $\mathcal{K}(R,(\nabla R)^2) = \mathcal{K}_1(R)-\mathcal{K}_2(R)(\nabla R)^2$. 
Under these assumptions,
the function $\Psi$ in Eq.\eqref{Psi-Naruko} can be expressed as
\begin{equation}
            \Psi(R,(\nabla R)^2,\Box R)= \mathcal{K}_1(R)-\mathcal{K}_2(R)(\nabla R)^2+ \mathcal{G}_1(R)\,\Box R, \label{Naruko-function}
\end{equation}
and the action \eqref{eq:Einstein-fBB0} becomes
\begin{eqnarray}
   S&=&  \int d^4x \sqrt{-\hat{g}}\Big[\frac{1}{2\kappa^2} \hat{R} -\frac{1}{2}\hat{g}^{\mu\nu} \nabla_\mu \chi\nabla_\nu \chi
   -\frac{1}{2\kappa^2}\hat{g}^{\mu\nu}e^{-\sqrt{\tfrac{2}{3}}\kappa\chi}\Big(\mathcal{G}_1'(\phi)+ \mathcal{K}_2(\phi)\Big)\nabla_\mu \phi\nabla_\nu \phi \nonumber\\
   && ~~~~~~~~~~~~~~~~-\frac{1}{2\kappa^2}e^{-\sqrt{\tfrac{2}{3}}\kappa\chi}\,\phi
   + \frac{1}{2\kappa^2}e^{-2\sqrt{\tfrac{2}{3}}\kappa\chi}\,\mathcal{K}_1(\phi)\Big].
   \label{eq:Einstein-KG}
\end{eqnarray}
Let us point out that, under the assumptions made here, the action given in Eq.(2.15) of \cite{Naruko2016} 
coincides with that in Eq.\eqref{eq:Einstein-KG} and, by properly choosing the functions $\mathcal{K}$ and $\mathcal{G}$, it can have healthy kinetic terms \cite{Naruko2016}.

Finally, introducing a new scalar degree of freedom $\sigma$ by means of an (invertible) field redefinition
\begin{equation}
\sigma=\mathcal{J}(\phi)\equiv\frac{1}{\kappa}\int \sqrt{\Big(\mathcal{G}_1'(\phi)+ \mathcal{K}_2(\phi)\Big)}d \phi+C,
\end{equation}
where $C$ is a constant, Eq.\eqref{eq:Einstein-KG} reduces to Einstein gravity coupled to two scalar fields, under the influence of the potential $\tilde W(\chi,\sigma)$, namely,
\begin{equation}
  S= \int d^4x \sqrt{-\hat{g}}\Big[\frac{1}{2\kappa^2} \hat{R} 
  -\frac{1}{2}\hat{g}^{\mu\nu} \nabla_\mu \chi\nabla_\nu \chi
  -\frac{1}{2}\hat{g}^{\mu\nu}e^{-\sqrt{\tfrac{2}{3}}\kappa\chi}\nabla_\mu \sigma\nabla_\nu \sigma 
  -\tilde{W}(\chi,\sigma)\Big],
  \label{eq:action-biscalar}
\end{equation}
with
\begin{equation}
   \tilde{W}(\chi,\sigma)=\frac{1}{2\kappa^2}e^{-\sqrt{\tfrac{2}{3}}\kappa\chi}\,
   \left(
   \mathcal{J}^{-1}(\sigma)-
   e^{-\sqrt{\tfrac{2}{3}}\kappa\chi}\,\mathcal{K}_1(\mathcal{J}^{-1}(\sigma))\right). 
   \label{Potenccial-Naruko-Final}
   \end{equation}

The field $\sigma$  can be expressed in terms of $R$ as follows:
\begin{equation}
 \sigma = \frac{1}{\kappa}\int \sqrt{
   \Big(\mathcal{G}_1'(R)+ \mathcal{K}_2(R)\Big)
 }\, dR+C.\label{function-calJ}
\end{equation}
Using Eq.\eqref{eq:lambda}, the scalar field $\chi$ in terms of $R$ is given by
\begin{equation}\label{chi-field}
 \chi = \frac{1}{\kappa} \sqrt{\frac{3}{2}}\, 
 \ln{\Bigg[ 
   \mathcal{K}_1'(R)
   + \mathcal{K}_2'(R)\,(\nabla R)^2
   + \Big(\mathcal{G}_1'(R) + 2\,\mathcal{K}_2(R)\Big)\Box R
   + \Box\mathcal{G}_1(R)
 \Bigg]}.
\end{equation}
We have therefore shown that the action in Eq.\eqref{action-Psi} under the assumptions leading to Eq.\eqref{Naruko-function} is equivalent, in the Einstein frame, to GR with two scalar degrees of freedom and a potential. The resulting bi-scalar action in Eq.\eqref{eq:action-biscalar} can be directly compared with that obtained in hybrid metric–Palatini gravity  in the scalar representation in the Einstein frame, which will be presented in the next section.

\section{The generalized hybrid gravity}
\label{sec:GH-gravity}

The action for the generalized hybrid metric-Palatini
gravity is given by
\begin{equation}
    S=\frac{1}{2\kappa^2}\int d^4x\sqrt{-g}~f\left(R,\mathcal{R}\right)+S_m,
    \label{action-hibrida}
\end{equation}
where $R=g^{\mu\nu}R_{\mu\nu}$ is the metric curvature scalar composed by the Levi-Civita connection $\Gamma^{\rho}_{\mu\nu}$, and $\mathcal{R}=g^{\mu\nu}\mathcal{R}_{\mu\nu}$ is the Palatini curvature scalar defined in terms of a torsionless
connection 
$\widetilde{\Gamma}_{\mu \nu}^\rho$ as $\mathcal{R}_{\mu\nu}=\widetilde{\Gamma}^{\lambda}_{~\mu\nu,\lambda}-\widetilde{\Gamma}^{\lambda}_{~\mu\lambda,\nu}+\widetilde{\Gamma}^{\sigma}_{~\mu\nu}\widetilde{\Gamma}^{\lambda}_{~\sigma\lambda}-\widetilde{\Gamma}^{\sigma}_{~\mu\lambda}\widetilde{\Gamma}^{\lambda}_{~\sigma\nu}$ 
which is assumed to be
independent of the metric
\cite{tamanini}. 

Variations of the action in Eq.\eqref{action-hibrida} with respect to the metric $g_{\mu\nu}$ and the independent connection $\widetilde{\Gamma}^\rho_{\mu\nu}$, lead to the following field equations 
\begin{equation}
    f_{,R}R_{\mu\nu}-\frac{1}{2}g_{\mu\nu}f-\left(\nabla_{\mu}\nabla_{\nu}-g_{\mu\nu}\Box_g\right)f_{,R}+f_{,\mathcal{R}}\mathcal{R}_{\mu\nu}=\kappa^2T_{\mu\nu},
    \label{field-equation}
\end{equation}
and
\begin{equation}
    \widetilde{\nabla}_{\rho}\left(\sqrt{-g}f_{,\mathcal{R}}g^{\mu\nu}\right)=0,
    \label{eq:3}
\end{equation}
where $\widetilde{\nabla}_{\mu}$ denotes the covariant derivative in terms of $\widetilde{\Gamma}^\rho_{\mu\nu}$ and $f_{,\mathcal{R}}\equiv \partial_\mathcal{R} f $. Eq.\eqref{eq:3} 
can be solved by defining $\sqrt{-\tilde{g}}\tilde{g}^{\mu\nu}=\sqrt{-g}f_{,\mathcal{R}}g^{\mu\nu}$, leading to
\begin{equation}
    \widetilde{\Gamma}_{\nu \lambda}^\mu=\frac{1}{2}\tilde{g}^{\mu\rho} (\partial_\nu \tilde{g}_{\rho \lambda}+\partial_\lambda \tilde{g}_{\rho \nu}-\partial_\rho \tilde{g}_{\nu \lambda} ) \label{gamma-palatini}.
\end{equation}
Hence, the connection 
$\widetilde{\Gamma}_{\nu \lambda}^\mu$ can be written as a Levi-Civita connection in terms of the conformal metric 
$\tilde{g}_{\mu\nu}=f_{,\mathcal{R}}g_{\mu\nu}$. Using Eq.\eqref{gamma-palatini}, the Ricci scalar associated
to the connection $\tilde\Gamma$ can be written as
\begin{equation}
\mathcal{R}_{\mu\nu}=R_{\mu\nu}+\frac{3}{2(f_{,\mathcal{R}})^2}\partial_\mu f_{,\mathcal{R}} \partial_\nu f_{,\mathcal{R}}-\frac{1}{ f_{,\mathcal{R}}}\left(\nabla_{\mu}\nabla_{\nu}+\frac{1}{2}g_{\mu\nu}\Box_g\right)f_{,\mathcal{R}}.
\label{ricci-calig}
\end{equation}
Taking the trace of Eqs.\eqref{ricci-calig} and \eqref{field-equation}, and rearranging them, we obtain the following differential equations for the curvature scalars:
\begin{equation}
\begin{aligned}
  &3\Box f_{,\mathcal{R}}- f_{,\mathcal{R}} R+ f_{,\mathcal{R}}\mathcal{R}-\frac{3}{2f_{,\mathcal{R}}}\partial_\mu f_{,\mathcal{R}} \partial^\mu f_{,\mathcal{R}}=0, \\
  &3\Box f_{,R}+ f_{,R} R+f_{,\mathcal{R}}\mathcal{R}-2f=\kappa^2 T. 
\end{aligned}
\label{dof-equations}
\end{equation}
Equations \eqref{dof-equations} show that, depending on the form of $f\left(R,\mathcal{R}\right)$, the GH theories may have
two additional degrees of freedom when compared to GR.
These equations, together with Eq.\eqref{field-equation}, compose the (highly coupled) system of differential equations to be solved in the GH theories. 

\subsection{Scalar-tensor representation}
To obtain the scalar-tensor representation of the GH theory, two auxiliary scalar fields $\alpha$ and $\beta$ are introduced such that the action in Eq.\eqref{action-hibrida} reads \cite{tamanini}
\begin{equation}\label{action-ab}
    S=\frac{1}{2\kappa^2}\int d^4x \sqrt{-g}\Big(f(\alpha,\beta)+f_{,\alpha}\ (R-\alpha)+f_{,\beta}\ (\mathcal{R}-\beta)\Big)+S_m,
\end{equation}
for $\alpha = R$ and $\beta = \mathcal{R}$, the action in Eq.\eqref{action-ab} reduces to the original action.

The variation with respect to $\alpha$ and $\beta$ gives the system
\begin{equation}
\begin{aligned}
f_{,\alpha\alpha}\ (R - \alpha) + f_{,\alpha\beta}\ (\mathcal{R} - \beta) &= 0, \\
f_{,\alpha\beta} \ (R - \alpha) + f_{,\beta\beta} \ (\mathcal{R} - \beta) &= 0,
\end{aligned}
\label{determinant-equations}
\end{equation}
can be interpreted as a linear homogeneous system in the variables $R - \alpha$ and $\mathcal{R} - \beta$, with coefficients given by the second derivatives of the function $f(\alpha, \beta)$. This system admits the trivial solution $    R = \alpha, \mathcal{R} = \beta$ provided that the determinant of the coefficient matrix is non-vanishing, \textit{i.e.},
\begin{equation}\label{constrain-hessiano}
f_{,\alpha\alpha} ~ f_{,\beta\beta} - \left(f_{,\alpha\beta} \right)^2 \neq 0.    
\end{equation}
Therefore, under this condition, the only solution is the one that identifies the scalar fields \(\alpha\) and \(\beta\) with the curvature scalars \(R\) and \(\mathcal{R}\), respectively. The constraint in Eq.\eqref{constrain-hessiano} excludes 
for instance 
linear functions
of $R$ and $\mathcal{R}$.

The action in Eq.\eqref{action-ab} can be rewritten as follows:
\begin{equation}
    S=\frac{1}{2\kappa^2}\int d^4x\sqrt{-g}\Big[ \varphi R -\psi \mathcal{R}-V(\varphi,\psi)\Big]+S_m,
\end{equation}
where  
\begin{equation}
    \varphi=f_{,\alpha}~~~~~~ \text{and} ~~~~~~\psi=-f_{,\beta}, \label{campos-Híbrida}
\end{equation}
The inclusion of the minus sign in the definition of $\psi$ ensures that the field does not acquire a negative kinetic energy. The interaction potential is defined as 
\begin{equation}
    V(\varphi,\psi)=-f(\alpha(\varphi),\beta(\psi))+\varphi \alpha(\varphi)-\psi\beta(\psi). \label{potencial-V-híbrida}
\end{equation}
Now, the first equation in \eqref{dof-equations} allows us to eliminate $\mathcal{R}$ from the action, yielding
\begin{equation}
    S=\frac{1}{2\kappa^2}\int d^4x\sqrt{-g}\left[ (\varphi-\psi) R-\frac{3}{2\psi}\partial_\mu \psi \partial^\mu \psi-V(\varphi,\psi)\right]+S_m.
\end{equation}
Defining a new scalar field 
$\vartheta$ by
$\vartheta=\varphi-\psi$, 
\begin{equation}
    S=\frac{1}{2\kappa^2}\int d^4x\sqrt{-g}\left[ \vartheta R-\frac{3}{2\psi}\partial_\mu \psi \partial^\mu \psi-W(\vartheta,\psi)\right]+S_m,
\end{equation}
where $W(\vartheta,\psi)\equiv V(\vartheta+\psi,\psi)$. 
At this point, we may transform the action to the Einstein frame by applying the conformal transformation $\hat{g}_{\mu\nu}=\vartheta g_{\mu\nu}$. We shall also  introduce the fields
\begin{equation}
    \chi = \sqrt{\frac{3}{2\kappa^2}}\,\ln \vartheta ~~~~ \text{and}~~~~\sigma= \frac{\sqrt{8}}{\kappa}\sqrt{\psi} .\label{campos-TH}
\end{equation}
The field $\chi$ is associated with the conformal factor, while $\sigma$ is needed to simplify the kinetic term of $\psi$. In terms of these variables, the action in the Einstein frame becomes
\cite{tamanini} 
\begin{equation}
    S=\int d^4x\sqrt{-\hat{g}}\left[\frac{1}{2\kappa^2} \hat{R} -\frac{1}{2}\hat{g}^{\sigma\nu} \nabla_\nu \chi\nabla_\sigma \chi-\frac{1}{2}\hat{g}^{\sigma\nu}e^{-\sqrt{\frac{2}{3}}\kappa\chi}\nabla_\sigma \sigma\nabla_\nu \sigma-\tilde{W}(\chi,\sigma)\right],
    \label{Rosa-action}
\end{equation}
where the Jordan-frame potential $W(\phi,\psi)$ is re-expressed in terms of the new scalar fields and yields the Einstein-frame potential $\tilde{W}(\chi,\sigma)$, given by
\begin{equation}
   \tilde{W}(\chi,\sigma)= \frac{1}{2\kappa^2}e^{-2\sqrt{\frac{2}{3}}\kappa \chi} V(e^{\sqrt{\frac{2}{3}}\kappa \chi}+\kappa^2\sigma^2/8,\kappa^2\sigma^2/8). \label{potencial-hibirda.scalar-tensor}
\end{equation}
The action in Eq.\eqref{Rosa-action} is 
very similar to the one in Eq.\eqref{eq:action-biscalar}, the only difference being in the dependence of the potential with the fields. As we shall see in the next section, this similarity provides a bridge between the $f(R,(\nabla R)^2,\Box R)$ theories defined by  Eq.\eqref{Naruko-function} and GH theories.

\section{Applications}
\label{sec:Applications}

\begin{figure}[H]
\centering
\begin{tikzpicture}[
    >=Latex,
    node distance=5cm,
    every node/.style={align=center}
]

\node (psi) [
    draw,
    thick,
    rounded corners=4pt,
    minimum width=3cm,
    minimum height=1cm
]
{$\Psi\!\left(R,(\nabla R)^2,\Box R\right)$};

\node (f) [
    right=4cm of psi,
    draw,
    thick,
    rounded corners=4pt,
    minimum width=3cm,
    minimum height=1cm
]
{$f(R,\mathcal{R})$};

\node (bottom) [below=1.4cm of $(psi)!0.5!(f)$] 
{$ \{a(t),\,\chi(t),\,\sigma(t),\tilde{W}\}$};

\node (top) [above=1.4cm of $(psi)!0.5!(f)$] 
{$\{\mathcal{K}_1(R),\,\mathcal{G}_1(R)\,\text{or}\,\mathcal{K}_2(R)\}$};

\draw[->, line width=1.1pt, bend left=30]
    (f) to node[midway, circle, draw, fill=white, inner sep=1.2pt] {$ii$} (psi);

\draw[->, line width=1.1pt, bend left=30]
    (psi) to node[midway, circle, draw, fill=white, inner sep=1.2pt] {$i$} (f);

\end{tikzpicture}

\caption{Illustration of the correspondence between the $\Psi$ and $f$ formulations.}
\label{fig:correspondence}
\end{figure}
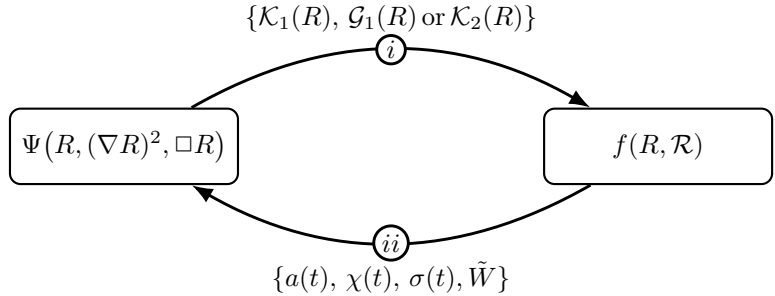

Before presenting illustrative examples that show how the correspondence developed in the previous section can be applied in a cosmological setting, let us give a summary of the method. There exist two possible paths to establish the bridge between the GH and GST theories. Figure~\ref{fig:correspondence} schematically summarizes these two procedures.

The first approach, corresponding to path~\textit{(i)} in Fig.~\ref{fig:correspondence}, consists of specifying the functions $\mathcal{K}_1(R)$, $\mathcal{G}_1(R)$, or $\mathcal{K}_2(R)$ that define the function $\Psi\!\left(R,(\nabla R)^2,\Box R\right)$, see Eq.\eqref{Naruko-function}. A given form of these functions
allows the construction of 
the corresponding scalar–tensor representation of the GH theory, with the scalar field $\sigma$ being related to the Ricci scalar through
Eq.\eqref{function-calJ}:
\begin{equation}\label{G1+Kyy2}
 \sigma = \mathcal{J}(R)= \frac{1}{\kappa}\int \sqrt{
   \Big(\mathcal{G}_1'(R)+ \mathcal{K}_2(R)\Big)
 }\, dR,
\end{equation}
and the scalar field $\chi$ being
a function of $R$  as in Eq.~\eqref{chi-field}:
\begin{equation}\label{chi-resume}
 \chi = \frac{1}{\kappa} \sqrt{\frac{3}{2}}\, 
 \ln{\Bigg[ 
   \mathcal{K}_1'(R)
   + \mathcal{K}_2'(R)\,(\nabla R)^2
   + \Big(\mathcal{G}_1'(R) + 2\,\mathcal{K}_2(R)\Big)\Box R
   + \Box\mathcal{G}_1(R)
 \Bigg]}.
\end{equation}
On the other hand, the potential defined in Eq.~\eqref{Potenccial-Naruko-Final} reads
\begin{equation} \label{potential-resume}
   \tilde{W}(\chi,\sigma)=\frac{1}{2\kappa^2}e^{-\sqrt{\tfrac{2}{3}}\kappa\chi}\,
   \left(
   \mathcal{J}^{-1}(\sigma)-
   e^{-\sqrt{\tfrac{2}{3}}\kappa\chi}\,\mathcal{K}_1(\mathcal{J}^{-1}(\sigma))\right).
\end{equation}
Next we take this potential as being 
equal to the one defined in the scalar-tensor representation of the GH theories in the Einstein frame, see Eq.\eqref{Rosa-action}, when rewritten in terms of the scalar fields
$\varphi$ and $\psi$ by means of Eq.~\eqref{campos-TH}. 
The relation between $\tilde W$ and the potential $V(\varphi,\psi)$
follows from 
Eq.~\eqref{potencial-hibirda.scalar-tensor}:
\begin{equation}\label{potencial-W-V}
      \frac{1}{2\kappa^2(\varphi-\psi)^2}
   V(\varphi,\psi)=\tilde{W}\!\left(
   \sqrt{\tfrac{3}{2\kappa^2}}\ln(\varphi-\psi),
   \tfrac{2\sqrt{2}}{\kappa}\sqrt{\psi}
   \right).
\end{equation}
The potential $V(\varphi,\psi)$ can then be related to the original hybrid function $f(R,\mathcal{R})$ and its derivatives. Substituting the relations \eqref{campos-Híbrida} into Eq.~\eqref{potencial-V-híbrida} directly yields

\begin{equation}\label{V-función-f}
 V\big(f_{,R},-f_{,\mathcal{R}}\big) = -f(R,\mathcal{R})
 + R f_{,R} + \mathcal{R} f_{,\mathcal{R}}.
\end{equation}
This equation is a
first-order partial differential equation of Clairaut type for
$f(R,\mathcal R)$. Hence, given the functions
$\mathcal{K}_1(R)$, $\mathcal{G}_1(R)$, or $\mathcal{K}_2(R)$,
one can reconstruct the associated hybrid function $f(R,\mathcal R)$. If the cosmological background $a(t)$ is known, the Ricci scalar $R(t)$ follows from its definition, and the scalar fields $\sigma(t)$ and $\chi(t)$ can be obtained explicitly as functions of time from Eqs.~\eqref{G1+Kyy2} and \eqref{chi-resume}, respectively.

\vspace{2mm}

The second approach, corresponding to path~\textit{(ii)} in Fig.~\ref{fig:correspondence}, starts from a given GH theory in the scalar–tensor representation and allows the reconstruction of the corresponding $\Psi\!\left(R,(\nabla R)^2,\Box R\right)$ formulation. In this approach, given the functions $\big(a(t), \chi(t), \sigma(t)\big)$ and the potential $\tilde{W}(\chi,\sigma)$, one can reconstruct $\Psi$ explicitly starting from relation~\eqref{function-calJ}, as follows. Since the Ricci scalar is determined by the scale factor through $R=R\big(a(t),\dot a(t),\ddot a(t)\big)$, and assuming $\dot R(t)\neq 0$, the relation $R(t)$ can be inverted to obtain $t=t(R)$. Equation~\eqref{function-calJ} can then be rewritten to yield the combination 
$\mathcal{G}_1'(R) + \mathcal{K}_2(R)$
\begin{equation}
    \kappa^2\frac{\dot{\sigma}^2\big(t(R)\big)}{\dot{R}^2\big(t(R)\big)}
    = \mathcal{G}_1'(R) + \mathcal{K}_2(R),
\end{equation}
where the auxiliary field $\sigma$ is defined in Eq.~\eqref{campos-TH}. 
Moreover, the potential in Eq.~\eqref{Potenccial-Naruko-Final}, when written in terms of $R$, satisfies
\begin{equation}\label{W(R)}
\tilde{W}\Big(\chi\!\big(t(R)\big), \sigma\!\big(t(R)\big)\Big)
= \frac{1}{2\kappa^2}\,R\,e^{-\sqrt{\frac{2\kappa^2}{3}}\,\chi\!\big(t(R)\big)} 
- \frac{1}{2\kappa^2}\,e^{-2\sqrt{\frac{2\kappa^2}{3}}\,\chi\!\big(t(R)\big)} 
\,\mathcal{K}_1\!\big(R\big).
\end{equation}
Therefore, it fixes, along with the fields, the functional dependence of $\mathcal{K}_1$ with $R$ for any background in which $R(t)$ is monotonic.

In the following subsections, we shall illustrate the two procedures with explicit examples, showing how to apply the correspondence for path~\textit{(i)} in the cases $\mathcal{G}_1(R)=0$ and $\mathcal{K}_2(R)=0$, and for path~\textit{(ii)} in a cosmological setting.

\subsection{\texorpdfstring{The $\Psi(R,(\nabla R)^2)$ Case}~}\label{subsec:X(R)nablaR}
Let us focus on a subclass of derivative-extended models of the form
\begin{equation}
\Psi(R,(\nabla R)^2)= \mathcal{K}_1(R)-\mathcal{K}_2(R)(\nabla R)^2, \label{X(R)(nabla R)}
\end{equation}
%
%
which corresponds to setting  $\mathcal{G}_1$ to zero 
in Eq.\eqref{Naruko-function}. They represent extensions of $f(R)$ gravity that incorporate a linear dependence on $(\nabla R)^2$ modulated by a function of $R$. This class of models has been explored in cosmological applications \cite{Chervon:2018, Chervon2019}, as well as black holes and wormholes \cite{Chervon2020b}, and spherically symmetric spacetimes 
\cite{Chervon-cuadratic}.

The scalar fields $\sigma$
and $\chi$, given by Eqs.
\eqref{G1+Kyy2} and \eqref{chi-resume}, can be written as follows:
\begin{equation}\label{integral-X}
    \sigma=\mathcal{J}(R)=\frac{1}{\kappa}\int\sqrt{\mathcal{K}_2(R)}dR,
\end{equation}
and
\begin{equation}
 \chi=\frac{1}{\kappa}\sqrt{\frac{3}{2}} \ln{\Big( \mathcal{K}_1'(R)
+\mathcal{K}_2'(R)\,(\nabla R)^2
+2\,\mathcal{K}_2(R)\Box R\Big)}.\label{campo-chi-X}
\end{equation}
The potential in Eq.\eqref{potential-resume}, which does not depend explicitly on the function $\mathcal{G}_1(\phi)$, takes the form
\begin{equation}
    \tilde{W}(\chi,\sigma)= \frac{1}{2\kappa^2}\mathcal{J}^{-1}(\sigma) e^{-\sqrt{\frac{2}{3}}\kappa\chi} -\frac{1}{2\kappa^2}e^{-2\sqrt{\frac{2}{3}}\kappa\chi}\mathcal{K}_1\big(\mathcal{J}^{-1}(\sigma)\big).\label{potential-X(R)}
\end{equation}
In what follows, we shall determine the function $f(R,\mathcal{R})$ 
that defined a GH theory 
for two representative cases of $\Psi$ in Eq.\eqref{X(R)(nabla R)}. The first corresponds to the model discussed in~\cite{Naruko2016}, which reduces to Einstein gravity with two minimally coupled scalar fields and is manifestly ghost-free. In the second 
case,  the function $\Psi$ is specifically chosen in such a way that 
a quasi-de Sitter scale factor
(see for instance \cite{Odintsov2021})
is a solution of the ensuing equations of motion.


\subsubsection{\texorpdfstring{The case $\mathcal{K}_2(R)=\tfrac{1}{2}$ and $\mathcal{K}_1(R)=0$. }~}
\label{sec:K2=1/2,K1=0}
In this case, the comparison with the GH theory in the scalar-tensor representation is straightforward, and the Einstein-frame action reduces exactly to
\begin{eqnarray}
  S&=&
   \int d^4x \sqrt{-\hat{g}}\Bigg[\frac{1}{2\kappa^2} \hat{R}
   -\frac{1}{2}\,\hat{g}^{\mu\nu}\nabla_\mu \chi\nabla_\nu \chi
   -\frac{1}{2}\,e^{-\sqrt{\frac{2}{3}}\kappa\chi}\,\hat{g}^{\mu\nu}\nabla_\mu \sigma\nabla_\nu \sigma
   -\frac{1}{\sqrt{2}\kappa}\,e^{-\sqrt{\frac{2}{3}}\kappa\chi}\,\sigma\Bigg],
\label{Action-Naruko-hibrida}
\end{eqnarray}
which corresponds to the potential
\begin{equation}
     \tilde{W}(\chi,\sigma)= \frac{1}{\sqrt{2}\kappa}\,\sigma\, e^{-\sqrt{\frac{2}{3}}\kappa \chi}.
\label{Potenccial-Naruko-hibrida}
\end{equation}
The scalar redefinitions that link the two-scalar Einstein-frame description to the curvature variable read
\begin{equation}
     \sigma=\frac{1}{\sqrt{2}\kappa}\,R + C,
     \qquad
     \chi= \frac{1}{\kappa}\sqrt{\frac{3}{2}}\ln{(\Box R)},
\end{equation}
To determine the corresponding $f(R,\mathcal{R})$ function, we start from the potential in Eq.\eqref{Potenccial-Naruko-hibrida} in the Einstein frame, and obtain $V(\varphi,\psi)$ through Eq.~\eqref{potencial-W-V}. It follows that
\begin{equation}
    \frac{1}{\sqrt{2}\kappa(\varphi-\psi)}\frac{2\sqrt{2}}{\kappa}\sqrt{\psi} 
    \;=\; \frac{1}{2\kappa^2(\varphi-\psi)^2} \, V(\varphi,\psi),
    \label{Relation-V-Naruko}
\end{equation}
which relates the Jordan-frame potential to the Einstein-frame potential \eqref{Potenccial-Naruko-hibrida}.  
Using Eq. \eqref{V-función-f}, the resulting differential equation for the function $f(R,\mathcal{R})$ can be written as
\begin{equation}
-4 \,(f_{,R}+f_{,\mathcal{R}})\sqrt{-f_{,\mathcal{R}}}
    + f_{,R}\,R + f_{,\mathcal{R}}\,\mathcal{R}-f=0.
    \label{caso-simple-naruko}
\end{equation}
This equation is a particular case 
of the Clairaut equation in two variables. Its general solution is linear in $R$ and $\mathcal{R}$, which implies that the resulting function $f$ does not satisfy the condition in Eq.\eqref{constrain-hessiano}. Therefore, only the singular solution of the Clairaut equation provides a consistent form of $f$ compatible with the required constraint. Thus, the acceptable  solution takes the form
\begin{equation}\label{sol->K1=0}
f(R,\mathcal{R})= \frac{1}{16}\,R^2\,(R-\mathcal{R}).
\end{equation}
This result shows that the particular theory of the form $\Psi\left(R,(\nabla R)^{2}\right)=-\tfrac{1}{2}(\nabla R)^{2}$ is dynamically equivalent to the specific GH theory given by Eq.\eqref{sol->K1=0}. Namely, any solution of the former is also a solution of the latter. 

\subsubsection{\texorpdfstring{The quasi-de Sitter Case}~}

As a second example, the function 
$f$ will be obtained from 
the functions 
$\mathcal{K}_1(R)$ and $\mathcal{K}_2(R)$ 
introduced in \cite{Chervon:2018}:
\begin{equation}\label{eq:example-f1-X}
  \mathcal{K}_1(R)=12\,R_*^2\,\epsilon\Big(2C_2 R_*\,e^{R/R_*}-C_1\Big),
  \qquad
  \mathcal{K}_2(R)=-C_1 - C_2\,e^{R/R_*}\big(R-2R_*+36\epsilon\big),
\end{equation}
where $C_1$ and $C_2$ are integration constants, and $R_*$ is a positive constant. The ensuing equations of motion in a cosmological setting
have the quasi-de Sitter scale factor 
\begin{equation}\label{quasi-a(t)}
  a(t)=\exp\!\Big(\sqrt{\tfrac{R_*}{12}}\,t-\epsilon\,t^2\Big),\qquad \epsilon\ll 1,
\end{equation}
as a solution
\cite{Chervon:2018}. 
%
As we shall see below, the explicit form of the potential \eqref{potential-X(R)} in terms of the fields $\varphi$ and $\psi$ will lead
to a differential equation for the function $f$. We now proceed to reconstruct the hybrid model $f(R,\mathcal{R})$ for the case $C_1=0$. From Eq.\eqref{integral-X}, it follows that
\begin{eqnarray}
\mathcal{J}\left(\phi\right)=\sqrt{C_2}e^{\frac{2R_*-3\phi-36\epsilon}{2 R_*}} \Big( e^{\frac{\phi}{R_*}} \left(2R_*-\phi-36\epsilon\right)\Big)^{3/2}E_{-\frac{1}{2}}\left[\frac{2R_*-\phi-36\epsilon}{2 R_*}\right]\label{eq:J-def}
\end{eqnarray}
where $E_n(z)$ denotes the generalized exponential integral
\footnote{ $E_n(z)=\int_1^\infty\ t^{-n}e^{-zt}dt$.}. 
The plots of $\mathcal{J}(\phi)$ for $\epsilon<<1$
show that $\mathcal{J}$ is approximately linear up to $\phi\approx1$. 
Keeping up to first order contributions in both $\epsilon$ and $\phi/R_*$, it follows that
\begin{align}
\label{eq:J-first-order}
\mathcal{J}^{(1)}(\phi)=
\sqrt{2C_2R_*}\,\Bigl[\,
\mathrm{e}\,\sqrt{\pi}\,\operatorname{erfc}(1)\,\bigl(R_* - 18\,\epsilon\bigr)
+ 2 R_* \Bigr]+ \sqrt{2C_2R_*}R_*\,\left(\frac{\phi}{R_*}\right)+\mathcal{O}\left(\tfrac{\phi^2}{R_*^2},\epsilon^2\right),
\end{align}
where
$\operatorname{erfc}(x)$ is the complementary error function and $\mathrm{e}$ is Euler's constant. 

\begin{figure}[H]
\centering
\includegraphics[scale=0.6]{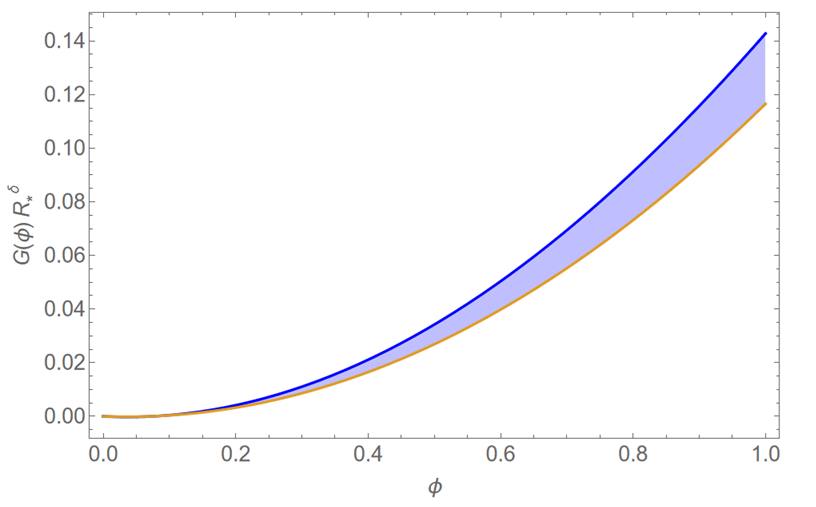}
\caption{Rescaled deviation $R_*^{\delta} G(\phi)$ with $\delta = 0.426$ for $R_* = 2$ (blue curve) and $R_* = 100$ (orange curve), and $\epsilon=10^{-3}$. The shaded region shows the range of $R_*^{\delta}G(\phi)$ as $R_*$ varies between $1$ and $100$. 
For $\phi \in [0,1]$, the magnitude is of order $10^{-1}$ in this domain. }\label{fig:diferencia}
\end{figure}
Fig.~\ref{fig:diferencia} shows the rescaled quantity 
$R_*^{\delta} G(\phi)$ with $\delta = 0.426$, and 
\begin{equation}
G(\phi) \equiv \mathcal{J}(\phi)
-\mathcal{J}^{(1)}(\phi).
\end{equation}
For $R_* \in [1,100]$ and $\phi \in [0,1]$, the deviation remains bounded between the curves corresponding to the extreme values $R_*=1$ and $R_*=100$, showing that the first-order expansion does not significantly deviate from the exact result in this regime.
Hence, $\mathcal{J}$
can be consistently inverted using $\mathcal{J}^{(1)}$. The inverse of the function $\mathcal{J}(\phi)$ can then be written as
\begin{equation}
     \mathcal{J}^{-1}(\sigma)=A_2\ \sigma+A_1+\mathcal{O}\big(\sigma^2,\epsilon^2\big)
\end{equation}
with the constants $A_1$ and $A_2$ defined by:
\begin{equation}
    A_1= -\Big( e \sqrt{\pi }\, \text{erfc}(1) (R_*-9 \epsilon )+2 (R_*+9 \epsilon )\Big), 
    \qquad  
    A_2= \frac{ 1}{\sqrt{2C_2R_*} }.
\end{equation}
Using this expression in the potential in Eq.\eqref{potential-X(R)}, it follows that

{\begin{equation}\label{potential-X(R)-final}
    \tilde{W}(\chi,\sigma)= \frac{1}{2\kappa^2}\big(A_1+A_2\,\sigma \big) e^{-\sqrt{\tfrac{2}{3}}\,\kappa\chi} 
    -\frac{1}{2\kappa^2}B_0\,e^{-2\sqrt{\tfrac{2}{3}}\,\kappa\chi}\left(1+\frac{A_2}{R_*}\sigma\right)+\mathcal{O}(\sigma^2,\epsilon^2),
\end{equation}}
where the constant $B_0$ is defined as follows:
\begin{equation}
    B_0=24 \epsilon R_*^3 C_2\,e^{\tfrac{A_1}{R_*}}.
\end{equation}
The potential in Eq.\eqref{potential-X(R)-final} is next used in relations \eqref{potencial-W-V} and \eqref{V-función-f} to determine the function $f(R,\mathcal{R})$. This procedure leads to the following differential equation for $f$:

{\begin{equation}\label{eq:f1-X}
    \big(A_1+A_2\,\sqrt{-8f_{,\mathcal{R}}} \big)\,(f_{,R}+f_{,\mathcal{R}})  
-B_0\left(1+\frac{A_2}{R_*}\sqrt{-8f_{,\mathcal{R}}}\right)
= -f+f_{,R}R+f_{,\mathcal{R}}\mathcal{R}.
\end{equation}}
Its singular solution takes the form
{\begin{equation}\label{Naruko a hibrida-ejemplo}
f(R,\mathcal{R}) =      \frac{1}{8 A_2^2}(A_1-R)^2 (R-\mathcal{R})+\frac{B_0}{R_*} (R-A_1)+B_0
\end{equation}}
The scalar fields as functions of $R$ are given by
\begin{equation}
     \sigma(R)={\mathcal{J}(R)},
\end{equation}
\begin{equation}
     \chi(R)=\ln{ \Bigg[C_2 e^{R/R_*}\!\left(
24\,\epsilon\,R_*^2
-\frac{R-R_*+36\epsilon}{R_*}\,(\nabla R)^2
-2\,\big(R-2R_*+36\epsilon\big)\,\Box R
\right) \Bigg]}.
\end{equation}
The Ricci scalar is given by
\begin{equation}
    R(t)= R_*-12\epsilon-8\epsilon\sqrt{3R_*}\: t+\mathcal{O}(\epsilon^2).
\end{equation}
Hence, starting from the quasi-de Sitter ansatz for the scale factor, the hybrid function $f(R,\mathcal{R})$ has been reconstructed together with the scalar fields $\sigma(t)$ and $\chi(t)$,
{in the regime $\left|\frac{{R}}{R_*}\right| \lesssim 1  $, with $\epsilon<<1$ . }

\subsection{\texorpdfstring{The $\Psi(R,\Box R)$ Case}~}\label{subsec:fRboxR}
In what follows, a subclass of derivative-extended models of the form
\begin{equation}\label{h+g-cuadritoR}
   \Psi(R,\Box R) = \mathcal{K}_1(R) + \mathcal{G}_1(R)\,\Box R,
\end{equation}
is considered. These models constitute the simplest nontrivial extension of $f(R)$ gravity, obtained by 
introducing a linear dependence on $\Box R$ modulated by a function of $R$, and 
have been studied in different contexts, such as inflation \cite{BERKIN1990348,Gott}, and avoidance of future singularities 
\cite{Carloni2018}.

In the case $\mathcal{G}_1(R)=\gamma R$, the scalar fields $\sigma$ (Eq.\eqref{G1+Kyy2}) and $\chi$ (Eq.\eqref{chi-resume}) take the form
\begin{equation}\label{sigma-gamma}
    \sigma=\frac{\sqrt{\gamma}}{\kappa}\,R + C,~~~~~~
\chi= \frac{1}{\kappa}\sqrt{\frac{3}{2}}\ln{\Big(\mathcal{K}_1'(R)+2\gamma\Box R\Big)}, 
\end{equation}
so that the potential \eqref{potential-resume}, which does not depend explicitly on the function $\mathcal{K}_2(\phi)$, takes the form
\begin{equation}
    \tilde{W}(\chi,\sigma)= 
    \frac{1}{2\kappa}\frac{\sigma}{\sqrt{\gamma}}
    e^{-\sqrt{\tfrac{2}{3}}\kappa\chi}
    - \frac{1}{2\kappa^2}
    e^{-2\sqrt{\tfrac{2}{3}}\kappa\chi}
    \,\mathcal{K}_1\!\left(\frac{\kappa\sigma}{\sqrt{\gamma}}\right). \label{potentail-gamma-R}
\end{equation}
The form of this potential coincides with the one previously obtained in models where $\mathcal{K}_1(R)$ is taken as a polynomial up to third order in $R$, namely $\mathcal{K}_1(R)=R+\alpha R^2+\beta R^3$
\cite{RomeroCastellanos:2018inv, Wands1994, BERKIN1990348}.

\vspace{0.2cm}

In order to establish the relation between the hybrid metric-Palatini theory $f(R,\mathcal{R})$ and the case $\Psi(R,\Box R)=\mathcal{K}_1(R)+\gamma R\,\Box R$, using Eq. \eqref{potencial-W-V} together with the potential \eqref{potentail-gamma-R}, we find that
\begin{eqnarray}
\frac{\sqrt{8} }{\sqrt{\gamma}}\sqrt{\psi}(\varphi-\psi)-\mathcal{K}_1\left(\sqrt{\frac{8}{\gamma}}\sqrt{\psi}\right)&=& V(\varphi,\psi).
\end{eqnarray}
Replacing the definitions of the scalar fields, namely $\varphi=f_{,R}$ and $\psi=-f_{,\mathcal{R}}$, into the above equation, we obtain
\begin{equation}
    \frac{\sqrt{8} }{\sqrt{\gamma}}\sqrt{-f_{,\mathcal{R}}}\,(f_{,R}+f_{,\mathcal{R}})
    -\mathcal{K}_1\left(\sqrt{\tfrac{8}{\gamma}}\sqrt{-f_{,\mathcal{R}}}\right)
    = -f(R,\mathcal{R})+f_{,R}R+f_{,\mathcal{R}}\mathcal{R}.
\end{equation}
Solving this differential equation yields the following explicit singular solution:
\begin{equation}\label{Naruko-Hibrida-eje2}
    f(R,\mathcal{R})= \mathcal{K}_1(R)+ \frac{\gamma}{8}\,R^2 \left(R-\mathcal{R}\right).
\end{equation}
This result shows that, at the level of the scalar--tensor representation, the model $\Psi(R,\Box R)=\mathcal{K}_1(R)+\gamma R\,\Box R$ can be reformulated as a specific HG model \eqref{Naruko-Hibrida-eje2}, establishing a direct mapping between both theories.

{Let us point out that the result in Eq.~\eqref{Naruko-Hibrida-eje2} coincides with that obtained in Eq.~\eqref{sol->K1=0} for the particular choice $\mathcal{K}_1(R)=0$. In this case, $\gamma=\tfrac{1}{2}$. Accordingly, the scalar fields in Eq.~\eqref{sigma-gamma}, together with the Einstein-frame potential $\tilde W$ given in Eq.~\eqref{potentail-gamma-R}, precisely match those presented in Sec.~\ref{sec:K2=1/2,K1=0}. This equivalence follows directly from the structure of the Einstein-frame 
action, since only the combination $\mathcal{G}_1'(R) + \mathcal{K}_2(R)$ enters the kinetic sector, as shown in Eq.~\eqref{eq:Einstein-KG}. Therefore, different decompositions of $\mathcal{G}_1(R)$ and $\mathcal{K}_2(R)$ that preserve this sum yield equivalent scalar--tensor representations and the same functional form of the hybrid theory.}

\subsection{\texorpdfstring{Cosmological solutions in $f(R,\mathcal{R})$}~}\label{subsec:cosmologicalSol}
As an example of the second
approach shown in Fig.\ref{fig:correspondence}, we shall show next that, starting from the analytical cosmological solutions obtained in \cite{Rosa2017} within the GH theories framework, the corresponding functions $\mathcal{K}_1(R)$, $\mathcal{K}_2(R)$, and $\mathcal{G}_1(R)$ that characterize the $\Psi(R,(\nabla R)^2,\Box R)$ action can be obtained, see Eq.\eqref{Naruko-function}. 
As shown in \cite{Rosa2017, Rosa.Thesis}, the scale factor for the matter dominated era, namely $ a(t)=a_{0}t^{2/3}$, is a solution of the equations of motion that follow from the scalar-tensor representation of the GH theory defined by 
\begin{equation}
f(R,\mathcal{R})=\frac{(R+\mathcal{R})^2}{16V_0}+\mathcal{B}(R-\mathcal{R}),
\label{f-potencial-V-híbrida}
\end{equation}
with the scalar fields given by
\begin{align}
    \varphi(t) = -\frac{\varphi_{0}}{a_{0}^{3}t} + \varphi_{1}, \label{Rosa-varphi}
\end{align}
and
\begin{align}
    \psi(t) = \left(-\frac{\psi_{0}}{2a_{0}^{3}t} + \frac{\psi_{1}}{2}\right)^{2}, \label{Rosa-psi}
\end{align}
where $\varphi_{0},\varphi_{1},\psi_{0},\psi_{1}$ are integration constants. 

In order for $f$ to satisfy condition~\eqref{constrain-hessiano}, 
the function $\mathcal{B}$ must be such that
\begin{equation}
\frac{d^2 \mathcal{B}(u)}{du^2} \neq 0, ~~~~~~~~~\text{where $u \equiv R - \mathcal{R}$}.
\end{equation}
The function in Eq.\eqref{f-potencial-V-híbrida} leads to a quadratic potential of the form
\begin{equation}
   V (\varphi,\psi) = V_{0}\,(\varphi-\psi)^{2}, \label{quadratic-potential-Rosa}
\end{equation}
where $V_{0}$ is a constant. 
Using the
expressions for $a(t)$, $\varphi(t)$, and $\psi(t)$
in the scalar–tensor potential given in Eq.\eqref{potencial-hibirda.scalar-tensor}, 
the quadratic potential in Eq.\eqref{quadratic-potential-Rosa} reduces to
\begin{equation}
    \tilde{W}(\chi,\sigma)=\frac{V_0}{2\kappa^2}. \label{quadratic-potential}
\end{equation}
Consequently, Eq.~\eqref{W(R)}, written in terms of the Ricci scalar, reads
\begin{equation}
    \frac{1}{2\kappa^{2}}\,R\,e^{-\sqrt{\frac{2\kappa^{2}}{3}}\,\chi(R)}
    \;-\;\frac{1}{2\kappa^{2}}\,e^{-2\sqrt{\frac{2\kappa^{2}}{3}}\,\chi(R)}\,\mathcal{K}_{1}(R)
    \;=\;\frac{V_{0}}{2\kappa^{2}}.
    \label{eq:K1condition}
\end{equation}
Expressing $\chi$ in terms of the scalars $\psi$ and $\varphi$ in Eq.\eqref{campos-TH}, and subsequently in terms of $R$, it follows that
\begin{equation}
    \mathcal{K}_{1}(R)
    = a_{1}+a_{2}R+a_{3}R^{2}+a_{4}R^{1/2}+a_{5}R^{3/2},
    \label{H-solution-pot-cuadratico}
\end{equation}
where $a_{1},a_{2},a_{3},a_{4}$ and $a_{5}$ are constants determined by the integration constants $\psi_{0},\psi_{1},\varphi_{1}$, as well as by the parameters $V_{0}$ and $a_{0}$ of the background solution
\footnote{Lagrangians involving non-integer powers of $R$
have also been obtained in metric $f(R)$ gravity, see for instance \cite{Numajiri2021,Sergio2023, Cui2024}).}.
Next, to reconstruct the functions $\mathcal{G}_1(R)$ and $\mathcal{K}_{2}(R)$, we make use of the relation~\eqref{function-calJ}. By inserting the explicit form of 
$\psi(t)$ (see \eqref{Rosa-psi}) in Eq.\eqref{campos-TH} and subsequently in Eq.\eqref{G1+Kyy2}, it follows that
\begin{equation}
\mathcal{G}_1'(R)+ \mathcal{K}_2(R)
= -\frac{1}{V_{0}}\,\frac{1}{R}.
\label{G-K2}
\end{equation}
Equation~\eqref{G-K2} constrains only the combination 
$\mathcal{G}_1'(R)+\mathcal{K}_2(R)$ rather than the individual functions. Therefore, different choices of  $\mathcal{G}_1$ and $\mathcal{K}_2$ that satisfy Eq.\eqref{G-K2} lead to dynamically equivalent theories, differing at most by a boundary term, as expected (see Sec. \ref{sec:EF-Representation of GST}). 
We shall present the case in which the theory is written 
solely in terms of $\Box R$, with no explicit $(\nabla R)^2$ 
contribution. Accordingly, we set
$\mathcal{K}_2(R)=0,$
and Eq.\eqref{G-K2} yields
\begin{equation}
\mathcal{G}_1(R)=\frac{1}{V_{0}}
\ln\!\left(\frac{c_{0}}{R}\right).
\end{equation}
The GST function then takes the form
\begin{equation}
\Psi\left(R,\Box R\right)= a_{1}+a_{2}R+a_{3}R^{2}
+a_{4}\sqrt{R}+a_{5}R^{3/2}+\frac{1}{V_{0}}
\ln\!\left(\frac{c_{0}}{R}\right)\Box R.
\end{equation}
Upon integration by parts, this theory is mapped into the 
$\mathcal{G}_1=0$ case, where the explicit 
$\Box R$ term is traded for an explicit $(\nabla R)^2$ contribution.

\section{Summary and Conclusions}
\label{sec:Conclusions}

In this work we have established a systematic correspondence between higher--derivative gravity theories $\Psi\left(R,(\nabla R)^2,\Box R\right)$ of the form given in Eq.\eqref{Naruko-function}
and generalized hybrid metric--Palatini models \(f(R,\mathcal{R})\). The correspondence 
follows from 
the scalar--tensor representation of both types of theories in the Einstein frame. By introducing auxiliary fields and performing conformal transformations, both classes reduce to Einstein gravity minimally coupled to two interacting scalar fields \((\chi,\sigma)\) with a nontrivial potential \(\tilde{W}(\chi,\sigma)\). 
In particular, we have shown, without using explicit solutions, that any GST theory of the form \eqref{Naruko-function}
defines a GH theory. This equivalence entails that 
any solution of the given GST theory is also a solution of the corresponding 
GH theory.
\vspace{0.2cm}

We have illustrated this correspondence
through explicit examples. In particular, the function $f(R,\mathcal{R})$, as well as 
 the scalar fields $\sigma(t)$ and $\chi(t)$ have been obtained for the case of kinetic extensions of $f(R)$ gravity, and 
for a GST theory defined by a $\Psi$ that has a  quasi--de Sitter \textit{Ansatz} for the scale factor as a solution.

 In addition, the analysis of models of the form $\mathcal{K}_1(R)+\gamma R\,\Box R$ shows that they are
 dynamically equivalent to a specific realization of hybrid metric--Palatini gravity. In this correspondence, the interaction term between the metric and Palatini curvature scalars acquires the quadratic structure $R^{2}(R-\mathcal{R})$, thereby establishing an explicit mapping between higher--derivative $\Psi(R,\Box R)$ theories and generalized hybrid frameworks.

The correspondence between the scalar--tensor representation of hybrid metric--Palatini gravity and the GST theories defined by Eq. \eqref{Naruko-function} makes it possible to reconstruct the effective functions $\mathcal{K}_{1}(R)$, $\mathcal{K}_{2}(R)$, and $\mathcal{G}_{1}(R)$ directly from known cosmological solutions specified by 
$\{a(t),\varphi(t),\psi(t)\}$ in the hybrid theory.
The resulting $\Psi$
has automatically the given expansion factor and the corresponding scalar fields as solutions of the ensuing equations of motion. For the case of a quadratic potential in the Jordan--frame representation of the hybrid theory, it is possible to explicitly reconstruct these functions, with the 
dust-like expansion law and the scalar field as input.

We have shown that the link between GST theories and GH theories obtained here is useful in a cosmological setting. Possible extensions of our results to scenarios such as black holes, wormholes, dynamical systems and weak-field limit will be dealt with in a future publication.

\printbibliography[heading=bibintoc]

\section*{Acknowledgments}

This study was financed in part by the
Coordenação de Aperfeiçoamento de Pessoal de Nível Superior - Brasil (CAPES) - Finance Code 001.

\end{document}